\documentclass[preprint,showpacs,preprintnumbers,amsmath,amssymb,superscriptaddress,10pt,graphicx]{revtex4}

\usepackage{graphicx}
\usepackage{dcolumn}
\usepackage{bm}
\usepackage{amssymb}

\newcommand{\bea}{\begin{eqnarray}}
\newcommand{\ena}{\end{eqnarray}}

\begin{document}

\title{Detecting relics of a thermal gravitational wave background in the early Universe}

\author{Wen Zhao}
\email{Wen.Zhao@astro.cf.ac.uk} \affiliation{School of Physics and
Astronomy, Cardiff University, Cardiff, CF24 3AA, United Kingdom}\affiliation{Wales Institute of Mathematical and Computational
Sciences, Swansea, SA2 8PP, United Kingdom} \affiliation{Department of Physics, Zhejiang University of Technology, Hangzhou, 310014, People's Republic of China}

\author{Deepak Baskaran}
\email{Deepak.Baskaran@astro.cf.ac.uk} \affiliation{School of
Physics and Astronomy, Cardiff University, Cardiff, CF24 3AA,
United Kingdom}\affiliation{Wales Institute of Mathematical and
Computational Sciences, Swansea, SA2 8PP, United Kingdom}

\author{Peter Coles}
\email{Peter.Coles@astro.cf.ac.uk} \affiliation{School of Physics and
Astronomy, Cardiff University, Cardiff, CF24 3AA, United Kingdom}

\date{\today}


\begin{abstract}

A thermal gravitational wave background can be produced in the early
Universe if a radiation dominated epoch precedes the usual
inflationary stage. This background provides a unique way to study
the initial state of the Universe. We  discuss the imprint of this
thermal spectra of gravitons on the cosmic microwave background
(CMB) power spectra, and its possible detection by  CMB
observations. Assuming the inflationary stage is a pure de Sitter
expansion we find that, if the number of e-folds of inflation is
smaller than $65$, the signal of this thermal spectrum  can be
detected by the observations of Planck and PolarBear experiments, or
the planned EPIC experiments. This bound can be even looser if
inflation-like stage is the sub-exponential.

\end{abstract}



\maketitle


\section{Introduction \label{section1}}

Understanding the expansion history of the Universe is a fundamental task of modern cosmology.
The current observations of the cosmic microwave background radiation (CMB) \cite{map5}, large scale structure \cite{sdss}, Type Ia supernova \cite{snia}, amongst others, have provided us with a relatively clear picture of the expansion history of the Universe since the photon decoupling at the redshift $z\sim10^3$. The history of the Universe prior to decoupling is currently primarily deduced through indirect evidences, such as the primordial abundances of the light elements. The primordial abundances hold the information about expansion history of the Universe at the epoch of nucleosynthesis corresponding to redshift $z\sim10^{10}$.

Going back even further, in order to solve the flatness, horizon and monopole problems in the standard hot big-bang cosmological model, various inflation-like scenarios  \cite{inflation} (or the so-called `kick' scenarios in \cite{kick}) for the expansionary history of the Universe at very high redshift have been proposed. These scenarios generically predict a nearly scale-invariance primordial power spectra of density perturbations and gravitational waves \cite{perturbation}, and have been indirectly supported by the observation of the large scale structure and CMB temperature and polarization anisotropies power spectra \cite{sdss,map5}.

In this Letter we shall analyze the possibility of observing imprints from a pre-inflationary stage. These imprints could give us a glimpse at the physical conditions of the very early Universe right to the time of its birth. It is natural to suppose that the Universe underwent a radiation dominated stage of expansion prior to the inflation-like acceleration phase \cite{thermal0}. Furthermore, it is reasonable to assume that, during this stage at temperatures higher than $\sim10^{19}~{\rm GeV}$ a thermal equilibrium between the various components, including gravitons, is maintained through gravitational interaction. In this scenario, as the Universe cooled down and the gravitons decoupled, a background of thermal relic gravitons with a black-body spectrum would be left behind \cite{kolb,thermal1}. This thermal background of gravitational waves will garner uncontaminated information about the thermal pre-inflationary period, and it's detection would give a unique chance to probe the physics of pre inflationary Universe inaccessible by other means.

As the Universe expands, and undergoes inflationary expansion, the gravitational waves would be strongly redshifted to very low frequencies \cite{review}. The thermal peak frequencies would correspond to the range probed by CMB experiments $\nu\sim 10^{-18}-10^{-15}~\rm{Hz}$. The gravitational wave background could leave an observable imprint in the temperature and polarization anisotropies of the CMB, which is expected to be detected by observations in the near future. In the current work, we shall discuss this signature in the CMB, and analyze the possibility of detecting this signature by the upcoming observations.

In Section \ref{section2} we shall briefly consider the main physical motivations for the existence of thermal gravitational wave background. We give simple estimates for the main characteristics of this field and relate them to inflationary parameters. In Section \ref{section3} we analyze the affect of this background on the CMB temperature and polarization anisotropies. Based on the WMAP 5 year data, we place upper bounds on conformal temperature (explained below). We compare the expected signal with the sensitivity of the upcoming CMB experiments, and study the feasibility of detecting this background. Finally, in Section \ref{section4}, we discuss the physical implications of an observable thermal gravitational wave background.

Throughout this Letter, we will work with units in which $c=\hbar=k_B=1$.


\section{Thermal background of relic gravitational waves \label{section2}}

Let us discuss the possibility of generating black-body spectra of gravitational waves at very high energy scales in the early Universe. We shall assume that the Universe was radiation dominated before the inflationary epoch, and that all the particle species were highly relativistic. The interaction rate for particles interacting solely through the gravitational force would be $\Gamma\simeq \mathcal{T}^5/M_{\rm pl}^4$, where $\mathcal{T}$ is the physical temperature in the Universe, and $M_{\rm pl}\equiv G^{-1/2}=1.22\times10^{19}{\rm GeV}$ is the Planck energy \cite{ZeldovichNovikov,kolb}. So long as this interaction rate is large in comparison with the expansion rate characterized by Hubble parameter $H\simeq \mathcal{T}^2/M_{\rm pl}$, the gravitons would remain in thermal equilibrium with other particles. However, this equilibrium will be violated once $\Gamma\lesssim H$.
At this time, the gravitons would decouple from the other particle species, leaving behind a free-streaming thermal graviton background. At the time when the gravitons decoupled, the interaction rate is $\Gamma\simeq H$, i.e. $\mathcal{T}\simeq M_{\rm pl}$. It should be mentioned that the conditions for a thermal equilibrium of gravitons with other particles existed only at very early times $\mathcal{T}>M_{\rm pl}$, where the physics is not properly understood. For this reason, the graviton background might not have a black-body spectrum, although it would be in a non-vacuum state. However, in this Letter, we shall restrict our considerations to a black-body spectrum as a concrete example of a non-vacuum state, similar to considerations in \cite{kolb,thermal1}. We shall assume that this thermal graviton background decoupled at temperature  $\mathcal{T}\simeq M_{\rm pl}$ from the surrounding matter.

\begin{figure}[t]
\centerline{\includegraphics[width=18cm,height=10cm]{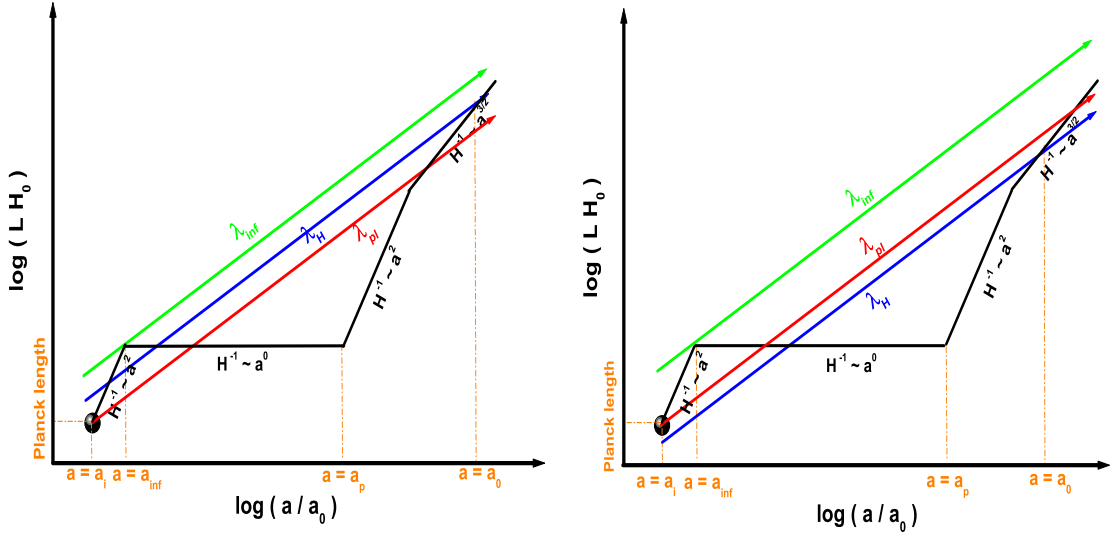}}
\caption{The evolution of Hubble radius, $H^{-1}$, and various physical sizes with the expansion of the Universe. The left panel shows the result in the scenario with $\lambda_{H}>\lambda_{pl}$, and   
 right panel shows the result in the scenario with $\lambda_{H}<\lambda_{pl}$.}\label{figure0}
\end{figure}

With the expansion of the Universe this graviton background would preserve its thermal spectrum, but would be strongly redshifted to very low temperatures. It is instructive to estimate the temperature of this background at the present epoch. In order to proceed, it is convenient to separate the history of the Universe into three stages: the initial radiation dominated stage, the inflationary stage and the post-inflationary stage. In terms of this division, the present day temperature $\mathcal{T}_0$ of the graviton background is
\bea
\label{temperature}
 \frac{\mathcal{T}_0}{M_{\rm
 pl}} \simeq\frac{a_i}{a_{inf}}\times\frac{a_{inf}}{a_{p}}\times\frac{a_{p}}{a_{0}},
\ena
where $a_i$, $a_{inf}$, $a_{p}$ and $a_{0}$ are the values of the scale factor at the time of graviton decoupling, the beginning of the inflationary stage, the end of inflation (beginning of the post-inflationary stage) and the present day, respectively. During and after the inflationary stage, the temperatures of the graviton field and that of the rest of the particle species behave in significantly different manner. The temperature on the graviton field strictly decreases with the expansion. On the other hand, at the end of inflation the temperature of the thermal bath, containing the rest of the particle species is significantly boosted by the process of reheating. Assuming that the observed CMB is the relic of this thermal bath, its temperature $\tilde{\mathcal{T}}_2$ at the beginning of the post-inflationary stage can be related to value of the scale factor at this stage through the relation $a_p/a_0=(2.73{\rm K}/\tilde{\mathcal{T}}_2)(3.91/106.75)^{1/3}$ \cite{kolb}. Using this expression, and denoting the temperature of the thermal bath at the beginning of the inflationary stage as $\tilde{\mathcal{T}}_1$, the equation (\ref{temperature}) can be rewritten as
\bea
\label{t0}
{\mathcal{T}_0}&\simeq& 8.0\times 10^{-27}(\tilde{\mathcal{T}}_1/\tilde{\mathcal{T}}_2)e^{60-N}{\rm K},
\ena
where $N \equiv \log(a_p/a_{inf})$ is number of e-folds during the inflationary stage. This spectrum is peaked at the frequency
\bea
\nu\simeq4.7\times10^{-16}(\mathcal{\tilde{T}}_1/\mathcal{\tilde{T}}_2)e^{60-N}{\rm
Hz}.
\ena
As can be seen, the peak frequency depends on the value of the number of e-folds during inflation, and the ratio of the temperatures $\tilde{\mathcal{T}}_1$ and $\tilde{\mathcal{T}}_2$. 

The number of e-folds $N$ cannot be very small. It must be larger than some $N_{\rm min}$ to account for the isotropy and homogeneity of the observed Universe. If we take the inflationary stage to be an exact de Sitter expansion at $\tilde{\mathcal{T}}_1\simeq 10^{16}{\rm GeV}$, a value $N_{\rm min}\simeq60$ follows (see for instance \cite{kinney,weinberg}).  In some realistic inflationary models, this minimum value could be even lower $N_{\rm min}\simeq 46$ \cite{kinney}. On the other hand, it is also important to consider the upper limits on the e-folds number. In general, the value of $N$ does not have a necessary upper bound. However, if we focus on specific inflationary models, the value of $N$ cannot be too large. For example, in order to account for the observed scalar spectral index $n_s=0.96$, in the inflationary model with the potential form $V(\phi)=\Lambda^4(\phi/\mu)^{p}$ with $p=2$, one has $N=50$. And  for the model with $p=4$, one has $N=74$ \cite{kinney}. In this section, as a rough estimation, we shall restrict our analysis to a typical range for the number e-folds $N\in(60,70)$.

In the case of an inflationary stage characterized by de Sitter expansion one has $\tilde{\mathcal{T}}_1\simeq\tilde{\mathcal{T}}_2$. Assuming $N\in(60,70)$, we get $\nu\in(2.1\times10^{-20},4.7\times10^{-16}) ~\rm{Hz}$. In the case of an inflationary stage with sub-de Sitter expansion, the value of $\tilde{\mathcal{T}}_1$ is typically larger than that of $\tilde{\mathcal{T}}_2$. Assuming $\tilde{\mathcal{T}}_1/\tilde{\mathcal{T}}_2=100$ and $N\in(60,70)$, we obtain $\nu\in(2.1\times10^{-18},4.7\times10^{-14})~\rm{Hz}$. It is important to note that in both the cases the peak frequency is in the range typically probed by the observations of the CMB. It is therefore reasonable to look for the signature of the relic thermal graviton spectrum in temperature and polarization anisotropies of the CMB. In the following section we shall address this question in more detail. However, it is also worth noting that, when the e-folds number is very large, $N\simeq100$ for example, the peak frequency will be $\nu\simeq10^{-30}$Hz, leaving no hope for observing the thermal nature of the graviton background. 

A natural question arises in the context of the evaluated above typical peak frequencies of the thermal graviton spectrum. Namely, is it possible for the peak wavelength $\lambda_{pl}$ of the thermal graviton spectrum to be larger than the minimum scale $\lambda_{inf}$ at which the Universe is homogenous and isotropic? The minimal scale of homogeneity and isotropy is set by the Grishchuk-Zel'dovich effect \cite{gz} at $\lambda_{inf}\sim 500\lambda_{H}$, where $\lambda_{H}=H_{0}^{-1}$ is the present day Hubble radius. If it were the case that $\lambda_{pl}>\lambda_{inf}$, the thermal nature of the graviton background would not be observable, since it would correspond to a peak frequency $\nu<10^{-21}$Hz, smaller than those accessible to CMB measurements. However, as illustrated in Fig. \ref{figure0}, such a situation does not arise. In Fig. \ref{figure0}, we plot the evolution of the Hubble radius $H^{-1}$ (black solid line) through the different stages of expansion of the Universe. During the radiation-dominated stage (before and after the inflationary stage), $H^{-1}\propto a^{2}$, and in the following matter-dominated era $H^{-1}\propto a^{3/2}$. In the inflationary stage, $H^{-1}$ is nearly a constant. The physical length associated with a length-scale of a fixed conformal length scales as $L\propto a$ with the expansion. In Fig. \ref{figure0} we show three such length scales: $\lambda_{pl}$ is the Planck length scale corresponding to the peak in the thermal graviton spectrum, $\lambda_{H}$ is the present day Hubble radius, and $\lambda_{inf}$ is the scale associated with large scale homogeneity and isotropy. Fig. \ref{figure0} shows that, if $\lambda_{inf}$ is taken to be equal to the Hubble length at the end of initial radiation dominated era at the beginning of the inflationary epoch, then it necessarily follows that $\lambda_{inf}>\lambda_{pl}$. This is a general statement, relying only on the assumption that the pre-inflationary state satisfies an equation of state $p/\rho>-2/3$. Thus, there are two possible scenarios. In the first case, when $\lambda_{H}>\lambda_{pl}$ (left panel in Fig. \ref{figure0}), corresponding to a small number of e-folds, the thermal nature of the graviton spectrum would be observable with CMB measurements. In the second case, $\lambda_{inf}>\lambda_{pl}>\lambda_{H}$ (right panel in Fig. \ref{figure0}), corresponding to a large number of e-folds, the thermal nature of the graviton spectrum would not be observable with CMB measurements.

Before proceeding to the analysis of the observable signature of a thermal background of gravitational waves in the anisotropies of the CMB, let us briefly analyze the resultant primordial power spectrum of this backround. The term {\it primordial power spectrum} denotes the power spectrum of relic gravitational waves in the radiation dominated epoch after the end of inflationary epoch when the wavelength of interest are significantly larger than the horizon. The thermal nature of the gravitational wave background prior to the inflationary stage will leave a distinct signature in the primordial power spectrum, which will then translate into a specific features in the power spectrum of CMB anisotropies. We start with the gravitational field of a slightly perturbed Friedmann-Lemaitre-Robertson-Walker universe given by
\bea
 ds^2=a^2(\eta)[-d\eta^2+(\delta_{ij}+h_{ij})dx^idx^j],
\ena
where $\eta$ is conformal time, $\delta_{ij}$ is the Kronecker delta symbol, and the metric perturbation field $h_{ij}$ only contains contribution from pure gravitational waves. The gravitational wave field has two modes of polarization, which can be expanded over spatial Fourier harmonics
\bea
 h_{ij}(\eta,{\bf x})&=&
 \frac{\sqrt{16\pi}}{a(\eta)M_{\rm  pl}}\int
 \frac{d^3{\bf k}}{(2\pi)^{3/2}} \times  \nonumber \\ &&
 \sum_{s=1,2} [c_{\bf k}f_k(\eta)+c_{-{\bf
 k}}^{\dag}f_k^*(\eta)]p_{ij}^{(s)}(\bf{k})e^{i{\bf k}\cdot{\bf x}}
 \nonumber \\
 &\equiv&\int\frac{d^3\bf{k}}{(2\pi)^{3/2}}  \sum_{s=1,2} h_{\bf k}(\eta)p_{ij}^{(s)}e^{i{\bf k}\cdot{\bf x}},
\ena
where $p_{ij}^{(s)}(\bf{k})$ is the polarization tensor \cite{zbg}.
The power spectrum is defined as
 \bea
 \langle h_{\bf k}h_{\bf
 k'}^*\rangle\equiv\frac{2\pi^2}{k^3}P_{t}(k)\delta^3({\bf k}-{\bf
 k'}),
 \ena
where the angle brackets indicate an ensemble average (see for example \cite{zbg}). Assuming vacuum initial conditions at a pre-inflationary stage, the Fourier coefficients satisfy the relations
\bea
\label{vacuuminitial}
\langle c_{\bf k}^{\dag}c_{\bf k'}\rangle=\delta^3({\bf k}-{\bf k'}),~~~
\langle c_{\bf k}c_{\bf k'}\rangle = \langle c_{\bf k}^{\dag}c_{\bf k'}^{\dag}\rangle = 0.
\ena
However, if we assume that the gravitational wave field was in thermal equilibrium, then the first relation in (\ref{vacuuminitial}) modifies to \cite{thermal1}
\bea
\label{blackbodyinitial}
\langle c_{\bf k}^{\dag}c_{\bf
k'}\rangle=\left(1+\frac{2}{e^{k/T}-1}\right)\delta^3({\bf k}-{\bf
k'}),
\ena
where $T$ is the conformal temperature of the gravitational wave background. The conformal temperature is related with the present day physical temperature $\mathcal{T}_0$ by the relation
\bea
\label{TandT0}
T={\mathcal{T}_0}a_0.
\ena
In the present Letter we set $a_0=1$, so that $T=\mathcal{T}_0 $.

The evolution of the universe through the inflationary expansion phase converts the initial vacuum into a multi-particle squeezed vacuum state characterized by a power law primordial spectrum \cite{rgw}. In the case of an initially thermal background, the primordial spectrum takes a modified form \cite{thermal1,thermal3}:
\bea
 P_t(k)=A_t(k_0)\left(\frac{k}{k_0}\right)^{n_t}\coth\left[\frac{k}{2T}\right],
\ena
where $n_t$ is the spectral index, which is close to zero for typical inflationary scenarios. $A_t(k_0)$ is the amplitude of the spectrum at the pivot wavenumber $k_0$. For large wavenumbers $k\gg T$, the power spectrum $P_t(k)\propto k^{n_t}$ is indistinguishable from the case of vacuum initial condition. However, for small wavenumbers $k\ll T$, the spectrum shows dissimilarity $P_t(k)\propto k^{n_t-1}$. Thus, generically, a thermal background exhibits difference in the power spectrum to the initial vacuum background at low wavenumbers $k<T$. This point was previously raised in \cite{grishchukreview}, were the author argued that modifications of $P_t$, due to an initial thermal spectrum, could be constrained by CMB observations. In the following discussion, we will also discuss the constraint on the conformal temperature $T$ by the 5-year WMAP observations.

\section{The imprint in the CMB and its determination\label{section3}}

\begin{figure}[t]
\centerline{\includegraphics[width=14cm,height=11cm]{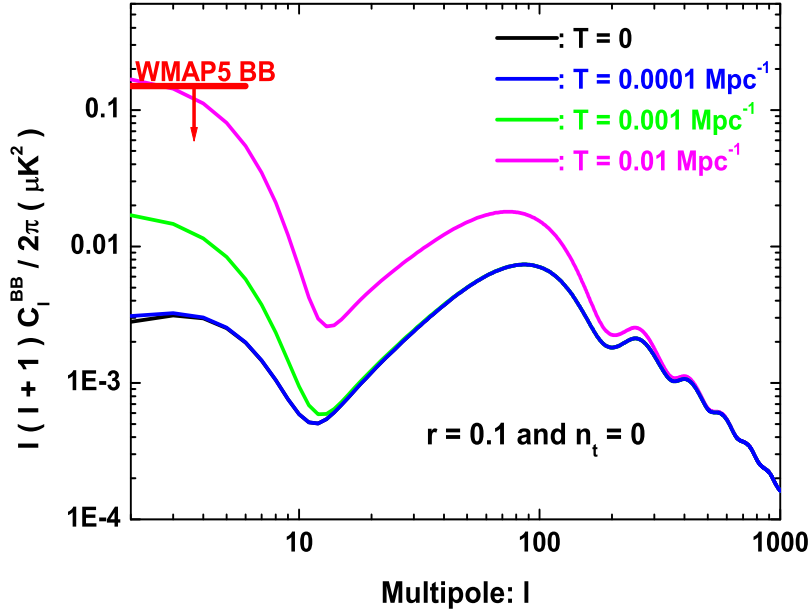}}
\caption{The $B$-mode polarization power spectrum $C_{\ell}^{BB}$ in the models with different conformal temperature $T$. In all the spectra $n_t=0$ and $r=0.1$ at the pivot wavenumber $k_0=0.2$Mpc$^{-1}$. The cosmological parameters are set at the WMAP5 best-fit value \cite{map5}.}\label{figure1}
\end{figure}

In this section we shall analyze the observable signatures of the thermal background of relic gravitational waves in the power spectrum of temperature and polarization anisotropies of the CMB. The power spectrum of CMB anisotropies due to gravitational waves are evaluated by solving the radiative transfer equation in the framework of perturbation theory \cite{polnarev,starobinskii,zaldarriaga} (see also \cite{gwcmb}). In general, gravitational waves leave their imprints in the four anisotropy spectra $C_{\ell}^{XX'} (XX'=TT,TE,EE,BB)$ \cite{zaldarriaga,gwcmb}. However, in this work, we shall restrict our discussion to the power spectrum of the so-called $B$-mode of polarization $C_{\ell}^{BB}$. The $B$-mode is solely generated (neglecting the possible foregrounds) by gravitational waves, and thus provides a clean channel for their detection. The features in the gravitational wave power spectrum $P_{t}$ at wavenumber $k$ translate predominantly into features in the CMB power spectrum at multipole $\ell\sim k\times10^{4}~\rm{Mpc}$ \cite{zb}. For this reason, we can hope to detect the signature of thermal gravitational wave background in the CMB for temperatures $T\gtrsim 0.0001$Mpc$^{-1}$. In Fig.~\ref{figure1}, we plot the CMB power spectrum $C_{\ell}^{BB}$ for various values of $T$. As expected, the signature of the thermal background is predominantly located at multipoles $\ell\lesssim T\times 10^{4}~\rm{Mpc}$.

\begin{figure}[t]
\centerline{\includegraphics[width=14cm,height=11cm]{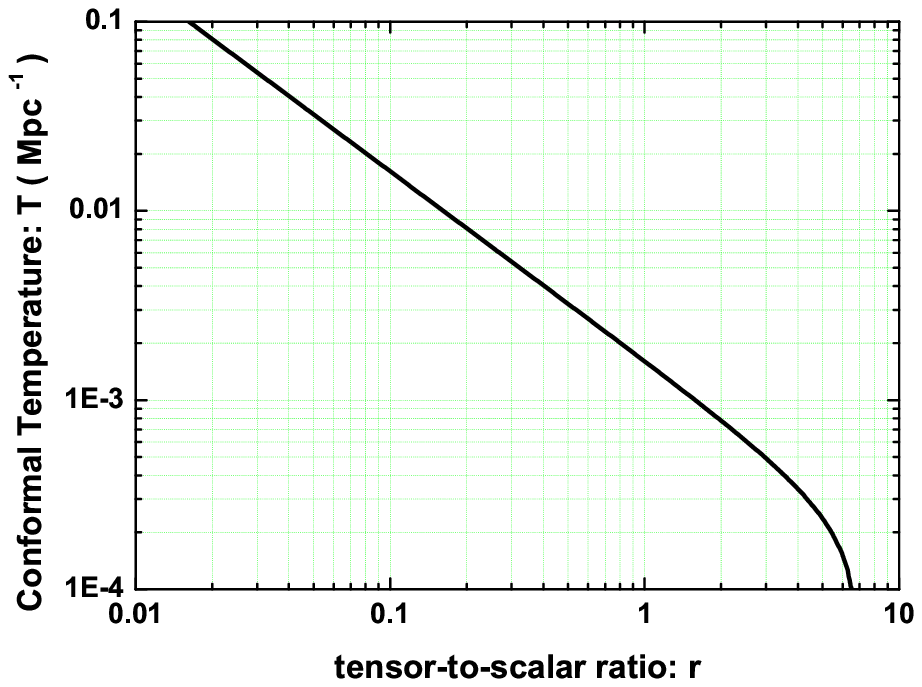}}
\caption{The upper limit of the conformal temperature $T$ as a function of the tensor-to-scalar ratio $r$, based on the WMAP5 observations.}\label{figure2}
\end{figure}

The current CMB observations have yet to detected the $B$-mode of polarization. The 5-year WMAP observation give an upper limit for the $B$-mode at lower multipoles, with the average value  $\ell(\ell+1) C_{\ell}^{BB}/2\pi$ at $\ell=2,3,4,5,6$ being smaller than $0.15\mu$K$^2$ ($95\%$C.L.)\cite{wmap5bb}. This upper limit, allows to place constraints on on the value of the conformal temperature depending on the tensor-to-scalar ratio $r$, shown in Fig.~\ref{figure2}. The tensor-to-scalar ratio $r$ characterizes the overall contribution of gravitational waves to the CMB anisotropies, and is defined as the ratio of the primordial power spectra of gravitational waves to density perturbations (see for example \cite{zb}). As expected, the constraints on $T$ scale inversely proportional to the value of $r$. For $r=0.1$ the constraints read $T<0.016$Mpc$^{-1}$, while for $r=0.3$ we have $T<0.005$Mpc$^{-1}$.

Let us now turn our attention to the analysis of prospects for future CMB observations. In order to determine the potential of the future CMB observations to constrain the conformal temperature of the thermal gravitational wave background we shall use an approach based on the Fisher information matrix. In terms of the Fisher matrix, the precision on a parameter $p_i$ that can potentially be attained is given by $\Delta p_{i}\simeq ({\bf F}^{-1})_{ii}^{1/2}$ \cite{fisher,fisher2}. In the present analysis we shall restrict to two free parameters $T$ and $r$.

The other cosmological parameters can be determined with high precision from the other CMB spectra $C_{\ell}^{TT}$, $C_{\ell}^{EE}$ and $C_{\ell}^{TE}$, and their inclusion in the analysis will not significantly alter the results. Therefore, we shall fix the other cosmological parameters at their fiducial values $\Omega_b=0.0456$, $\Omega_c=0.228$, $\Omega_{\Lambda}=0.726$, $\Omega_{k}=0$, $h=0.705$ and $n_s=0.96$, $A_s=2.036\times10^{-9}$ at the pivot wavenumber $k_0=0.2$Mpc$^{-1}$ \cite{map5}. For simplicity, in this Letter, we shall also fix the gravitational wave spectral index $n_t=0$, corresponding to a flat (scale-invariant) power spectrum.

The Fisher matrix can be written as
\cite{fisher}
\bea
\label{fishermatrix}
 {\rm {\bf F}}_{ij}=\sum_{\ell}\frac{\partial C_{\ell}^{BB}}{\partial p_{i}}
 \frac{\partial C_{\ell}^{BB}}{\partial p_{j}}\frac{1}{(\Delta
 D_{\ell}^{BB})^2},
\ena
where $i,j=1,2$ correspond to the parameters $T$ and $r$ respectively. The quantity $\Delta D_{\ell}^{BB}$ is the standard deviation of the estimator $D_{\ell}^{BB}$ \cite{zbg}
\bea
\Delta D_{\ell}^{BB}=\sqrt{\frac{2}{(2\ell+1)f_{\rm sky}}} (C_{\ell}^{BB}+N_{\ell}^{BB}).
\ena
In the above expression the noise power spectrum $N_{\ell}^{BB}$ and the sky cut factor $f_{\rm sky}$ are determined by the specificities of the particular CMB experiment, see for example \cite{zbg}.

We now estimate the potential to constrain the conformal temperature $T$ based on combining the data from the space-based Planck satellite \cite{planck} and the ground-based PolarBear experiment \cite{polarbear}. The former is sensitive to the $B$-mode of polarization at $\ell\lesssim20$, while the latter one is sensitive at $\ell>20$. Their combination provides an excellent opportunity to determine the gravitational wave signal \cite{zz}. The corresponding instrumental noises and the sky cut factors can be found in \cite{planck,polarbear}. In addition to the instrumental noises, in calculating the noise power spectrum  $N_{\ell}^{BB}$, we have included the contribution due to the cosmic lensing effect \cite{lensing}.

\begin{figure}[t]
\centerline{\includegraphics[width=14cm,height=11cm]{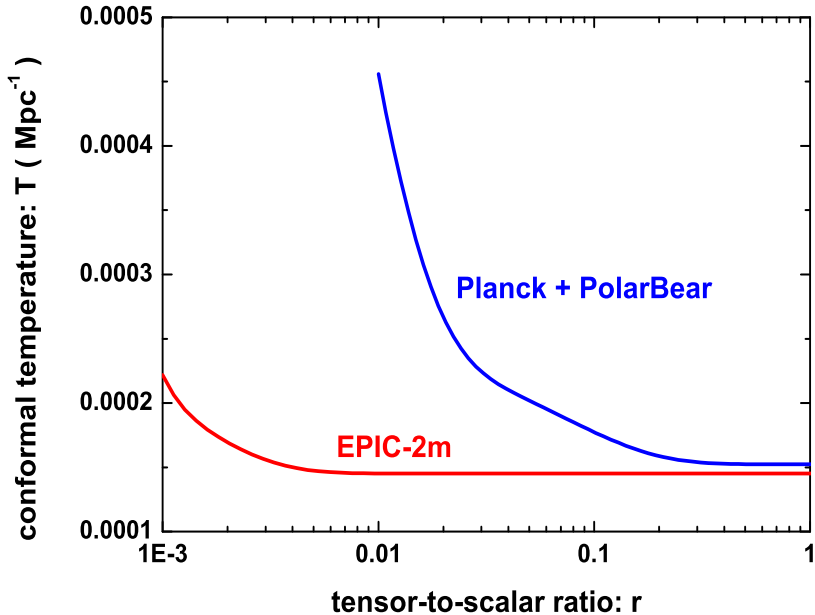}}
\caption{The curves show the smallest value of conformal temperature $T$ which can be determined by CMB experiments. The blue line shows the sensitivity of Planck$+$Polarbear experiment,  and the red line shows the sensitivity of the proposed EPIC-2m experiment.}
\label{figure3}
\end{figure}

Using expression (\ref{fishermatrix}), we calculate the value of $\Delta T$ for given values of parameters $T$ and $r$. For a fixed value of $r$, the smallest measurable value of $T$ is determined by the condition $T=\Delta T$. This lowest bound on the detectable signal is shown by blue line on Fig. \ref{figure3}. For a a typical value $r=0.1$, the attainable limit on the conformal temperature is $T=1.8\times10^{-4}$Mpc$^{-1}$. The lower limit on the detectable signal could be further improved by the proposed EPIC-2m experiment \cite{cmbpol} shown with the red line on Fig. \ref{figure3}. In this case, the lower limit is $T=1.4\times10^{-4}$Mpc$^{-1}$ for values $r\gtrsim 0.005$.

\section{Discussion and conclusions\label{section4}}

The possible determination or placement of upper limits on the conformal temperature of the thermal gravitational wave background would allow to place interesting constraints on the physics of the inflationary era. To see this, let us rewrite Eq.~(\ref{t0}) using Eq.~(\ref{TandT0}) in the form
\bea
\label{last}
T=0.017\times e^{60-N}(\tilde{\mathcal{T}}_1/\tilde{\mathcal{T}}_2){\rm Mpc}^{-1}.
\ena
Assuming a fiducial value $r=0.1$, the 5-year WMAP data places an upper limit on the conformal temperature $T<0.016$ (see Fig.~\ref{figure2}). Using Eq.~(\ref{last}), this limit translates into a constraint on the parameters of the inflationary expansion $N>60+\log(\tilde{\mathcal{T}}_1/\tilde{\mathcal{T}}_2)$. Assuming $\tilde{\mathcal{T}}_1\simeq\tilde{\mathcal{T}}_2$, we obtain a constraint on the number of e-folds $N\gtrsim 60$. It is worth pointing out that, this bound is consistent with the e-folds parameter required to solve the flatness, horizon and monopole problems in the standard hot big-bang cosmological model \cite{weinberg}.

The EPIC-2m experiment would be able to determine the conformal temperature for a broad range of tensor-to-scalar ratios $r\gtrsim 0.005$ down to a limit $T\gtrsim 1.4\times10^{-4}$Mpc$^{-1}$ (see Fig.~\ref{figure3}). A positive detection of non-zero conformal temperature by the EPIC-2m experiment would place upper bounds on the number of e-folds $N<65+\log(\tilde{\mathcal{T}}_1/\tilde{\mathcal{T}}_2)$, which would depend on the ratio of background temperatures before and after inflation $\tilde{\mathcal{T}}_1/\tilde{\mathcal{T}}_2$. For $\tilde{\mathcal{T}}_1\simeq\tilde{\mathcal{T}}_2$ the upper bound would be $N\lesssim65$, while for $\tilde{\mathcal{T}}_1 \simeq 100 \tilde{\mathcal{T}}_2$ the upper bound would become $N\lesssim69$.

The above considerations show that an observable thermal gravitational wave background may allow to place stringent constraints on the range of viable inflationary models. More over, its detection may shed light onto quantum gravity effects, which become important at Planck energy scales. On the other hand, an absence of observational indications of a thermal background would indicate one of the two possibilities. Either the initial state of the gravitational wave background was not thermal, or alternatively, that the number of e-folds was $N\gtrsim 69$ so that the present day conformal temperature is redshifted to $T\lesssim1.4\times10^{-4}$Mpc$^{-1}$.

Finally, it is worth pointing out that, along with gravitational waves, a thermal spectrum of density perturbations may have also existed in the very early Universe \cite{fang,warm,thermal2}. However, the nature of this spectrum depends crucially on the content and state of the matter in the very early Universe, questions which are yet to be fully understood. In addition, the evolution of the spectrum depends strongly on the physical specificities of inflation, which are also not fully understood. For this reason, without further simplifying assumptions, density perturbations cannot be directly used to probe the initial state of the Universe.


~

{\it Acknowledgements:}
The authors appreciate stimulating discussions with
L.P.Grishchuk, Y.Zhang and  K.Bhattacharya. W.Zhao is partially
supported by NSFC grants No.10703005 and No.10775119.


\end{document}